%% file: main.tex
\documentclass[manuscript, nonacm]{acmart}

\author{Qi Zhao}
\email{qiz1@umbc.edu}
\author{Marjory Pineda}
\email{mpineda1@umbc.edu}
\author{Ketul Chhaya}
\email{kchhaya1@umbc.edu}
\affiliation{%
  \institution{University of Maryland, Baltimore County}
  \state{Maryland}
  \country{USA}
}

\author{Aakash Gautam}
\email{aakash@pitt.edu}
\affiliation{%
  \institution{University of Pittsburgh}
  \state{Pennsylvania}
  \country{USA}
}

\author{Yasmine Kotturi}
\email{kotturi@umbc.edu}
\affiliation{%
  \institution{University of Maryland, Baltimore County}
  \state{Maryland}
  \country{USA}
}

\authorsaddresses{}
\setcopyright{none}
\makeatletter
\newcommand{\forceprintacmref}{%
  \vspace{-1.5em}
  \bgroup
  \par\noindent{\small\bfseries ACM Reference Format:}\par\nobreak\noindent
  \bgroup
    \def\\{\unskip{}, \ignorespaces}%
    {\small\authors}%
  \egroup
  {\small. \@acmYear. \@title\ifx\@subtitle\@empty\else: \@subtitle\fi. }%
  {\small In \textit{Proceedings of ACM CI 2026 and ACM HCOMP 2026 Posters and Demos (CI ’26 and HCOMP ’26)}. ACM, Alexandria, VA, USA, 3 pages.}%
  \egroup
}
\makeatother

\makeatletter
\AtBeginDocument{%
  \let\orig@ps@firstpagestyle\ps@firstpagestyle
  \def\ps@firstpagestyle{%
    \orig@ps@firstpagestyle
    \fancyfoot[L]{}
    \fancyfoot[R]{\thepage}
    \fancyfoot[C]{}
  }
  
  \let\orig@ps@standardpagestyle\ps@standardpagestyle
  \def\ps@standardpagestyle{%
    \orig@ps@standardpagestyle
    \fancyfoot[LO,RE]{}
    \fancyfoot[LE,RO]{\thepage}
    \fancyfoot[C]{}
  }
}
\makeatother

\begin{document}

\title{Evaluating Beyond the Screen: Collective Assessment of AI-Generated Business Plans with Resource-Constrained Entrepreneurs}

\begin{abstract}
\input{Sections/0-abstract}
\end{abstract}

\maketitle
\forceprintacmref
\input{Sections/1-introduction}
\input{Sections/2-system-methods}

\bibliographystyle{ACM-Reference-Format}
\bibliography{main}

\appendix

\end{document}

%% file: Sections/0-abstract.tex
Entrepreneurs increasingly use end-user generative AI technologies such as ChatGPT for high-stakes documents like loan applications and business plans, where AI-generated errors---a wrong price, a fabricated product---can affect loan or funding outcomes.
Current approaches to supporting evaluation of AI-generated text assume a single user assessing output alone, on screen. 
This can be especially demanding for resource-constrained entrepreneurs, whose digital and AI skills vary widely.
In this early-stage work, we explore how evaluation might instead be organized in a group setting and completed as a collective activity.
We extended BizChat, an AI-powered business-planning tool, with an evaluation module that links each generated claim to the entrepreneur's original input.
We partner with community organizations in Maryland---embedding BizChat within various entrepreneurship programs---where workshop attendees (N=14) evaluated their plans through think-pair-share discussion.
Early findings suggest interface scaffolds like claim-to-input links primed attendees with concrete, personal evaluations, which the group setting then extended beyond the screen: attendees requested printed copies, used rubrics to compare across plans, and drew on peers' knowledge to verify what they could not easily judge alone.

%% file: Sections/1-introduction.tex
\vspace{-10pt}
\section{Introduction and Related Work}
Entrepreneurs increasingly use end-user generative AI technologies such as ChatGPT for high-stakes tasks, such as managing finances \cite{schloskyChatGPTFinancialAdvisor2025}, completing loan applications \cite{saiGenerativeAIFinance2025}, and writing business plans~\cite{ashtariHumanAICollaborationSocial}.
These documents must reflect the specifics of their business such as products, pricing, and operations to secure funding, and AI-generated errors---a wrong price, a fabricated product---can affect loan or funding outcomes~\cite{bergerMoreCompleteConceptual2006, wuUnleashingPowerText2025}.
Current approaches to evaluating AI-generated text prioritize automated fact-checking~\cite{krishnaGenAuditFixingFactual2025}, benchmark comparisons~\cite{wangBenchmarkSelfEvolvingMultiAgent2025}, crowdsourced ratings~\cite{deSupernotesDrivingConsensus2024}, and, auditing frameworks~\cite{vadlamaniAuditingAISystems2026}. 
Across these approaches, however, evaluation is framed as a solo, on-screen task, where even human-evaluation studies rely on recruited participants completing assigned tasks~\cite{jungHumanAICollaborationLarge2026} rather than stakeholders evaluating documents they depend on.
This framing is especially demanding for resource-constrained entrepreneurs~\cite{zhangExploratoryBricolageHow2024}, who often lack access to professional feedback on business documents~\cite{dawaEntrepreneurialBricolageEntrepreneurial2026} and whose digital and AI skills vary widely~\cite{otisUnevenImpactGenerative2024}.
It also overlooks a resource these entrepreneurs already have: each other.
Entrepreneurs in resource-constrained communities already lean on peers and local organizations for business support: low-tech, in-person community collectives sustained entrepreneurs' digital engagement~\cite{huiCommunityCollectivesLowtech2020,kotturi2022tech}, and digital literacy itself can be measured as a capacity communities hold collectively rather than individually~\cite{dillahuntDevelopmentNewMeasure2025}.
Evaluating AI-generated text, too, might be organized as a collective activity, particularly for entrepreneurs least served by individualized tools.

%% file: Sections/2-system-methods.tex
\vspace{-10pt}
\section{The BizChat Evaluation Module: A Probe for Collective Evaluation}
\begin{figure}
    \centering
    \includegraphics[height=2.8in]{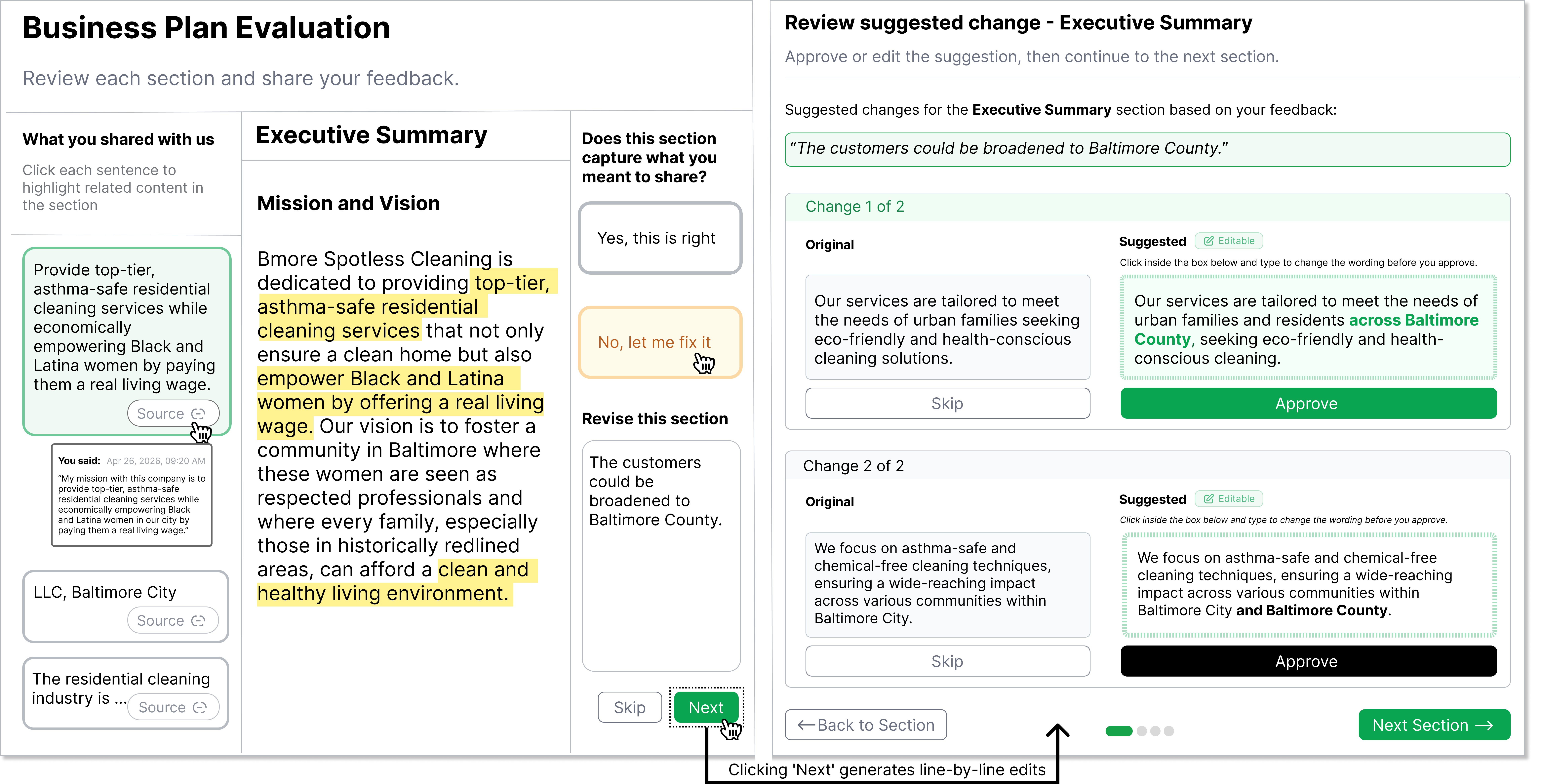}
    \caption{The BizChat evaluation module. (Left) The primary evaluation interface displays each generated section beside what the entrepreneur shared during onboarding; clicking a source card highlights the corresponding generated text in the business plan. (Right) The revision interface allows users to review, edit, and approve specific AI-suggested changes based on their feedback.}
    \Description{A three-panel interface: the entrepreneur's original statements on the left, the generated business-plan section with source-linked highlighted text in the center, and rating and revision controls on the right.}
    \label{fig:evluation}
    \vspace{-12pt}
\end{figure}

In this early-stage work, we explore how the evaluation of AI-generated text might be organized---and assessed---as a collective activity. 
As one step towards this collective activity, we first extended our system BizChat\footnote{BizChat is available to use at \url{https://bizchat.dev}}~\cite{romerolauroBizChatScaffoldingAIPowered2025a, romerolauroDesigningResilienceCommunityCentered2026}, an AI-powered business-planning web application, with an evaluation module that helps entrepreneurs verify whether their generated plan accurately reflects what they described. 
This module design draws on prior work showing that fine-grained attribution---linking AI-generated text to specific user inputs---reduces the verification burden on users~\cite{slobodkinAttributeFirstThen2024}.
For each plan section (e.g., Executive Summary, Service Line), the interface displays three panels side-by-side (Figure~\ref{fig:evluation} Left)---what the entrepreneur originally said alongside how the system summarized it, the generated plan text, and a feedback section for users' ratings and revisions.
Users trace whether their input was faithfully carried through by reviewing each section in this structured format (Figure~\ref{fig:evluation} Right); these interactions make evaluation tractable for a single user, providing critical scaffolding for the collective evaluation as described in the next section.

\vspace{-10pt}
\section{Approach and Early Findings: Supporting and Assessing Collective Evaluation}
Following a community-based participatory research (CBPR) approach \cite{israelREVIEWCOMMUNITYBASEDRESEARCH1998a}, we are partnering with local organizations\footnote{Community partners include the Housing Authority of the City of Annapolis, Baltimore Community Lending, UMBC Alex. Brown Center for Entrepreneurship} to embed BizChat within ongoing community programming---to date, we have conducted three sessions on AI and entrepreneurship with 14 workshop attendees.
In each, after hands-on BizChat use, attendees discussed their evaluations together (such as through Think-Pair-Share activities \cite{lymanResponsiveClassroomDiscussion1981} with prompts like ``How did you assess or trust what BizChat gave you?'')
The module scaffolded the collective evaluation by making verification shareable: because each claim was linked back to the entrepreneur's own words, checking whether the plan reflected what they had said became a visible comparison a neighbor could inspect and weigh in on.
Paper rubrics extended this shared context, giving attendees common criteria for evaluating one another's plans.
Follow-up interviews with two attendees suggest evaluation extended beyond the workshop itself: both took printed copies of their plans, one shared hers with two people for feedback---one peer ``said it is official,'' and another ``wanted to bulk it up''.
Building on these sessions, we plan to continue the series, alternating individually-focused and group-based evaluation activities to further investigate the role of peer discussion and collective assessment.
Over time, we aim to assess evaluation as a community-level capacity, drawing on emerging measures of collective digital literacy \cite{dillahuntDevelopmentNewMeasure2025} and on log data from BizChat's in-the-wild deployment.

%% file: main.bib
@String{Computing = "Computing" }

@String{Computer = "{IEEE} Computer" }

@article{israelREVIEWCOMMUNITYBASEDRESEARCH1998a,
    title = {{REVIEW} {OF} {COMMUNITY}-{BASED} {RESEARCH}: {Assessing} {Partnership} {Approaches} to {Improve} {Public} {Health}},
    volume = {19},
    issn = {0163-7525, 1545-2093},
    shorttitle = {{REVIEW} {OF} {COMMUNITY}-{BASED} {RESEARCH}},
    url = {https://www.annualreviews.org/content/journals/10.1146/annurev.publhealth.19.1.173},
    doi = {10.1146/annurev.publhealth.19.1.173},
    language = {en},
    number = {Volume 19, 1998},
    urldate = {2025-10-02},
    journal = {Annual Review of Public Health},
    publisher = {Annual Reviews},
    author = {Israel, Barbara A. and Schulz, Amy J. and Parker, Edith A. and Becker, Adam B.},
    month = may,
    year = {1998},
    pages = {173--202},
}

@misc{krishnaGenAuditFixingFactual2025,
    title = {{GenAudit}: {Fixing} {Factual} {Errors} in {Language} {Model} {Outputs} with {Evidence}},
    shorttitle = {{GenAudit}},
    url = {http://arxiv.org/abs/2402.12566},
    doi = {10.48550/arXiv.2402.12566},
    urldate = {2025-11-18},
    publisher = {arXiv},
    author = {Krishna, Kundan and Ramprasad, Sanjana and Gupta, Prakhar and Wallace, Byron C. and Lipton, Zachary C. and Bigham, Jeffrey P.},
    month = jan,
    year = {2025},
    note = {arXiv:2402.12566 [cs]},
}

@misc{jungHumanAICollaborationLarge2026,
    title = {Human-{AI} {Collaboration} in {Large} {Language} {Model}-{Integrated} {Building} {Energy} {Management} {Systems}: {The} {Role} of {User} {Domain} {Knowledge} and {AI} {Literacy}},
    shorttitle = {Human-{AI} {Collaboration} in {Large} {Language} {Model}-{Integrated} {Building} {Energy} {Management} {Systems}},
    url = {http://arxiv.org/abs/2602.16140},
    doi = {10.48550/arXiv.2602.16140},
    urldate = {2026-03-04},
    publisher = {arXiv},
    author = {Jung, Wooyoung and Jeon, Kahyun and Babon-Ayeng, Prosper},
    month = feb,
    year = {2026},
    note = {arXiv:2602.16140 [cs]
version: 1},
}

@inproceedings{romerolauroDesigningResilienceCommunityCentered2026,
    address = {Barcelona Spain},
    title = {Towards {Designing} for {Resilience}: {Community}-{Centered} {Deployment} of an {AI} {Business} {Planning} {Tool} in a {Pittsburgh} {Small} {Business} {Center}},
    isbn = {979-8-4007-2278-3},
    shorttitle = {Towards {Designing} for {Resilience}},
    url = {https://dl.acm.org/doi/10.1145/3772318.3791654},
    doi = {10.1145/3772318.3791654},
    language = {en},
    urldate = {2026-04-20},
    booktitle = {Proceedings of the 2026 {CHI} {Conference} on {Human} {Factors} in {Computing} {Systems}},
    publisher = {ACM},
    author = {Romero Lauro, Quentin and Gautam, Aakash and Kotturi, Yasmine},
    month = apr,
    year = {2026},
    pages = {1--19},
}

@inproceedings{huiCommunityCollectivesLowtech2020,
    address = {Honolulu HI USA},
    title = {Community {Collectives}: {Low}-tech {Social} {Support} for {Digitally}-{Engaged} {Entrepreneurship}},
    isbn = {978-1-4503-6708-0},
    shorttitle = {Community {Collectives}},
    url = {https://dl.acm.org/doi/10.1145/3313831.3376363},
    doi = {10.1145/3313831.3376363},
    language = {en},
    urldate = {2024-11-23},
    booktitle = {Proceedings of the 2020 {CHI} {Conference} on {Human} {Factors} in {Computing} {Systems}},
    publisher = {ACM},
    author = {Hui, Julie and Barber, Nefer Ra and Casey, Wendy and Cleage, Suzanne and Dolley, Danny C. and Worthy, Frances and Toyama, Kentaro and Dillahunt, Tawanna R.},
    month = apr,
    year = {2020},
    pages = {1--15},
}

@incollection{lymanResponsiveClassroomDiscussion1981,
    title = {The Responsive Classroom Discussion},
    booktitle = {Mainstreaming Digest},
    author = {Lyman, Frank},
    editor = {Anderson, A. S.},
    year = {1981},
    publisher = {University of Maryland College of Education}
}

@misc{vadlamaniAuditingAISystems2026,
    title = {Towards {Auditing} {AI} {Systems} in the {Wild}},
    url = {http://arxiv.org/abs/2606.17367},
    doi = {10.48550/arXiv.2606.17367},
    urldate = {2026-07-20},
    publisher = {arXiv},
    author = {Vadlamani, Aditya T. and Srinivasan, Anutam and Parthasarathy, Srinivasan},
    month = jun,
    year = {2026},
    note = {arXiv:2606.17367 [cs.CY]},
}

@inproceedings{romerolauroBizChatScaffoldingAIPowered2025a,
    address = {Amsterdam Netherlands},
    title = {{BizChat}: {Scaffolding} {AI}-{Powered} {Business} {Planning} for {Small} {Business} {Owners} {Across} {Digital} {Skill} {Levels}},
    isbn = {979-8-4007-1397-2},
    shorttitle = {{BizChat}},
    url = {https://dl.acm.org/doi/10.1145/3707640.3731928},
    doi = {10.1145/3707640.3731928},
    language = {en},
    urldate = {2026-03-04},
    booktitle = {Adjunct {Proceedings} of the 4th {Annual} {Symposium} on {Human}-{Computer} {Interaction} for {Work}},
    publisher = {ACM},
    author = {Romero Lauro, Quentin and Gautam, Aakash and Kotturi, Yasmine},
    month = jun,
    year = {2025},
    pages = {1--4},
}

@inproceedings{wangBenchmarkSelfEvolvingMultiAgent2025,
    address = {Abu Dhabi, UAE},
    title = {Benchmark {Self}-{Evolving}: {A} {Multi}-{Agent} {Framework} for {Dynamic} {LLM} {Evaluation}},
    shorttitle = {Benchmark {Self}-{Evolving}},
    url = {https://aclanthology.org/2025.coling-main.223/},
    urldate = {2026-07-21},
    booktitle = {Proceedings of the 31st {International} {Conference} on {Computational} {Linguistics}},
    publisher = {Association for Computational Linguistics},
    author = {Wang, Siyuan and Long, Zhuohan and Fan, Zhihao and Huang, Xuanjing and Wei, Zhongyu},
    editor = {Rambow, Owen and Wanner, Leo and Apidianaki, Marianna and Al-Khalifa, Hend and Eugenio, Barbara Di and Schockaert, Steven},
    month = jan,
    year = {2025},
    pages = {3310--3328},
}

@misc{deSupernotesDrivingConsensus2024,
    title = {Supernotes: {Driving} {Consensus} in {Crowd}-{Sourced} {Fact}-{Checking}},
    shorttitle = {Supernotes},
    url = {http://arxiv.org/abs/2411.06116},
    doi = {10.48550/arXiv.2411.06116},
    urldate = {2026-07-21},
    publisher = {arXiv},
    author = {De, Soham and Bakker, Michiel A. and Baxter, Jay and Saveski, Martin},
    month = nov,
    year = {2024},
    note = {arXiv:2411.06116 [cs.SI]},
}

@article{dillahuntDevelopmentNewMeasure2025,
  title   = {The Development of a New Measure of Collective Digital Literacy: Community Digital Capacity},
  author  = {Dillahunt, Tawanna R. and Shedden, Kerby and Filipof, Mila Ekaterina and Lee, Soyoung and Naseem, Mustafa and Toyama, Kentaro and Hui, Julie},
  journal = {Proceedings of the ACM on Human-Computer Interaction},
  volume  = {9},
  number  = {CSCW},
  year    = {2025},
  publisher = {ACM}
}

@misc{otisUnevenImpactGenerative2024,
    address = {Rochester, NY},
    type = {{SSRN} {Scholarly} {Paper}},
    title = {The {Uneven} {Impact} of {Generative} {AI} on {Entrepreneurial} {Performance}},
    url = {https://papers.ssrn.com/abstract=4671369},
    doi = {10.2139/ssrn.4671369},
    language = {en},
    urldate = {2026-07-22},
    publisher = {Social Science Research Network},
    author = {Otis, Nicholas and Clarke, Rowan and Delecourt, Solène and Holtz, David and Koning, Rembrand},
    month = feb,
    year = {2024},
}

@article{dawaEntrepreneurialBricolageEntrepreneurial2026,
    title = {Entrepreneurial bricolage and entrepreneurial success among female entrepreneurs in resource-constrained settings: a qualitative study},
    volume = {38},
    issn = {0827-6331, 2169-2610},
    shorttitle = {Entrepreneurial bricolage and entrepreneurial success among female entrepreneurs in resource-constrained settings},
    url = {https://www.tandfonline.com/doi/full/10.1080/08276331.2025.2536944},
    doi = {10.1080/08276331.2025.2536944},
    language = {en},
    number = {3},
    urldate = {2026-07-22},
    journal = {Journal of Small Business \& Entrepreneurship},
    author = {Dawa, Samuel and Mulira, Fiona and Aruo, Francis},
    month = may,
    year = {2026},
    pages = {441--470},
}

@inproceedings{kotturi2022tech,
  title={Tech help desk: Support for local entrepreneurs addressing the Long Tail of computing challenges},
  author={Kotturi, Yasmine and Johnson, Herman T and Skirpan, Michael and Fox, Sarah E and Bigham, Jeffrey P and Pavel, Amy},
  booktitle={Proceedings of the 2022 CHI Conference on Human Factors in Computing Systems},
  pages={1--15},
  year={2022}
}

@inproceedings{zhangExploratoryBricolageHow2024,
    title = {Exploratory {Bricolage}: {How} {Resource}-constrained {Entrepreneurs} {Achieve} {Stretch} {Goals}},
    shorttitle = {Exploratory {Bricolage}},
    url = {https://research.cbs.dk/en/publications/exploratory-bricolage-how-resource-constrained-entrepreneurs-achi/},
    doi = {10.5465/AMPROC.2024.15734abstract},
    language = {English},
    urldate = {2026-07-22},
    booktitle = {Proceedings of the {Eighty}-fourth {Annual} {Meeting} of the {Academy} of {Management}},
    publisher = {Academy of Management},
    author = {Zhang, Wei and Wang, Liyan and Li, Peter Ping},
    year = {2024},
}

@article{ashtariHumanAICollaborationSocial,
    title = {Human-{AI} {Collaboration} for {Social} {Entrepreneurs}: {AI}-{Driven} {Business} {Plan} {Evolution} {Aligned} {With} {Sustainable} {Development} {Goals}},
    language = {en},
    author = {Ashtari, Shiva and Mandegari, Siavash and Das, Ashok K and Allen, Janet K and Mistree, Farrokh},
}
